\documentclass[11pt,a4paper,twocolumn,]{article}
\usepackage{aas_macros}
\usepackage{graphics,graphicx}
\usepackage[usenames,dvipsnames]{color}
\usepackage{xspace}
\usepackage{natbib}
\usepackage{amsmath,amssymb,latexsym}

\usepackage[compact]{titlesec}

\titleformat*{\section}{\large \scshape}
\titlespacing*{\section}{0pt}{1.0\baselineskip}{0.0pt}

\titleformat*{\subsection}{\large \itshape}
\titlespacing*{\subsection}{0pt}{0.9\baselineskip}{0.0pt}

\titleformat*{\subsubsection}{\normalsize \slshape}

\usepackage{setspace}
\newcommand{\la}{\lesssim}
\newcommand{\micron}{$\mu$m~}
\newcommand{\arcsec}{$''$\xspace}

\newcommand{\eten}[1]{$10^{#1}$}

\newcommand{\kms}{\textrm{km~s}\ensuremath{^{-1}}\xspace}	

\newcommand{\macc}{\ensuremath{\dot{M}_{\mathrm {acc}}}\xspace}	
	
\newcommand{\mout}{\ensuremath{\dot{M}_{\mathrm{out}}}\xspace}

\newcommand{\ha}{\ensuremath{\textrm{H}\alpha}\xspace}

\newcommand{\pab}{\ensuremath{\textrm{Pa}\beta}\xspace}
\newcommand{\brg}{\ensuremath{\textrm{Br}\gamma}\xspace}

\newcommand{\herschel}{{\it Herschel}\xspace}

\newcommand{\mm}{millimeter\xspace}
\newcommand\submm{submillimeter\xspace}

\newcommand{\fir}{far-IR\xspace}

\newcommand{\fco}{{$F_{\rm CO}$}\xspace} 
\newcommand{\lmech}{{$L_{\rm mech}$}\xspace}

\begin{document}

\twocolumn[
  \begin{@twocolumnfalse}

\begin{center}
{\Large{\sc
{Interstellar medium and star formation studies with the Square Kilometre Array}
}}
\end{center}
\vspace*{-0.2cm}
\centerline{\sc
{P. Manoj$^1$, S. Vig$^2$, Maheswar, G.$^3$, U. S. Kamath$^4$ and  A. Tej$^2$}}

\begin{center}
{\footnotesize
$^1$ {Tata Institute of Fundamental Research, Homi Bhabha Rd, Mumbai 400 005~~(manoj.puravankara@tifr.res.in)}\\
$^2$ { Indian Institute of Space Science and Technology, Valiamala, Thiruvananthapuram 695 547} \\ 
$^3$ { Aryabhatta Research Institute of Observational Sciences, Manora Peak, Nainital, 263 129} \\ 
$^4$ { Indian Institute of Astrophysics, Sarjapur Road, Koramangala, Bangalore 560034}\\
}
\smallskip
{\it (To appear in Journal of Astrophysics and Astronomy (JOAA) special issue on "Science with the SKA: an Indian perspective")}
\end{center}
\vspace*{1pt}

\begin{abstract}
\noindent
Stars and planetary systems are formed out of molecular clouds in the interstellar medium. Although the sequence of steps involved in star formation are generally known, a comprehensive theory which describes the details of the processes that drive formation of stars is still missing. The Square Kilometre Array (SKA), with its  unprecedented sensitivity and angular resolution, will play a major role in filling these gaps in our understanding.  In this article, we present a few science cases that the Indian star formation community is interested in pursuing with SKA, which include investigation of AU-sized structures in the neutral ISM, the origin of thermal and non-thermal radio jets from protostars and the accretion history of protostars, and formation of massive stars and their effect on the surrounding medium.

\end{abstract}

\bigskip


\vspace*{1.0cm}

\end{@twocolumnfalse}
 ]

\section{Introduction}

The formation of stars and planetary systems out of interstellar clouds is one of the central problems in contemporary astrophysics.  Multi-wavelength observational studies, augmented by theoretical and laboratory studies in the last three decades or so have been successful in providing a framework to understand the formation of stars and planetary systems.  In the current paradigm for low-mass star formation, the process begins with the gravitational collapse of a slowly rotating cloud core, leading to the formation of a central protostar surrounded by a rotating disk and an overlying envelope from which the material rains down onto the disk \citep[see reviews:][]{shu87, mo07}. Since the total angular momentum is conserved during the collapse, the high angular momentum material in the outer envelope first collapses to form a disk, before getting accreted onto the central protostar \citep{ulrich76,cm81,tsc84}. In the early embedded stages, the system drives powerful bipolar jets/outflows, the origin of which is not entirely understood. As the system evolves, the envelope dissipates either by draining onto the disk or is cleared out by stellar winds and outflows, leaving behind a young pre-main sequence star surrounded by a disk.  Planetary systems are formed out of such protoplanetary disks which are the natural byproducts of star formation process \citep[e.g.][]{sp05,hart09}.   Although the different stages in the formation of star and planetary systems are broadly understood, the details of the various processes that drive star and planet formation are only poorly understood.  Several questions remain to be answered. How are molecular clouds formed out of the neutral Interstellar medium (ISM) and how long do they last? What processes controls the efficiency of star formation in molecular clouds? What determines the final stellar mass? What drives the accretion in protostars?  What is the launching mechanisms for the jets in young stars? How are they collimated and accelerated to such large distances? In addition, as opposed to low-mass star formation, there is no coherent picture yet for the formation of massive stars. 

The Square Kilometre Array (SKA), with its  unprecedented sensitivity and angular resolution, will play a major role in answering several of these questions.  Below we discuss a few of problems in interstellar medium and star formation that SKA will directly address.

\section{Investigation of the AU-sized structures in ISM: SKA-TMT Synergies}
The 21-cm absorption observable in the spectra of bright continuum sources in the background are used to probe the properties of the cold atomic component of interstellar medium (ISM). Several interferometric studies of the bright extended sources have revealed the existence of structures in cold HI gas \citep{1962ApJ...135..151C, 1965ApJ...142.1398C}. Subsequently, over the past few decades, both observations and theory have provided ample evidence for the existence of structures in the ISM on scales from $\sim$1 kpc down to $\sim$1 pc \citep[e.g., ][]{1990ARA&A..28..215D}. Adopting typical values of thermal pressure, temperature and observed column density of the cold neutral medium (CNM) as $P_{th}\sim2250$ cm$^{-3}$ K \citep{2001ApJS..137..297J}, T$\sim$70 K \citep{2003ApJ...586.1067H} and $5\times10^{19}$ cm$^{-2}$ \citep{2003ApJ...586.1067H}, the typical expected scale length for CNM feature is about 1 pc. Thus it came as a surprise when observers began to report structures on AU scales in several sightlines. These results were based on spatial mapping of the HI absorption-line profiles across extended extragalactic background sources \citep[e.g.,][]{1976ApJ...206L.113D, 1989ApJ...347..302D, 1996MNRAS.283.1105D, 2001AJ....121.2706F, 2005AJ....130..698B, 2008MNRAS.388..165G, 2012ApJ...749..144R}, temporal and spatial variations of optical interstellar absorption lines (like NaI D and CaIIK) against binary stars \citep{1996ApJ...464L.179M, 1996ApJ...473L.127W, 2008MNRAS.388..323W, 2013MNRAS.429..939S}, globular clusters \citep{1999ApJ...520L.103M, 2009PASP..121..606W}, and the time variability of HI absorption profiles against high proper motion pulsars \citep{1992MNRAS.258P..19D, 1994ApJ...436..144F, 2003MNRAS.341..941J}. In addition to these, the presence of AU-sized structures in the ionized \citep{1987Natur.326..675F, 1987Natur.328..324R, 2007ASPC..365..299W} and the molecular \citep{1993ApJ...419L.101M, 2000A&A...355..333L} components of ISM, though not as prevalent as in CNM, has invoked further interest in the topic.

The thermal pressure calculated for the observed AU-sized structures, assuming them as blobs of HI gas having a spherical geometry, is found to be much higher than the hydrostatic equilibrium pressure of the ISM or the standard thermal pressure of the CNM \citep{1997ApJ...481..193H}. Thus it is difficult to comprehend the existence of these structures in pressure equilibrium with other components of ISM. Also, such over-dense and over-pressured structures are expected to be short-lived yet are omnipresent in observations. Several explanations were proposed to reconcile the observations. \cite{1997ApJ...481..193H} suggested that the observed AU-sized structures are actually gas distribution in nonspherical geometries like curved filaments and/or sheets that happen to be aligned along the line of sight. On the other hand, \cite{2000MNRAS.317..199D} suggested that the observations are basically statistical fluctuations resulting from compact, overdense gaseous structures in interstellar medium that exist in a wide range of spatial scales.The scintillation phenomenon combined with the velocity gradient across the absorbing H I gas was suggested by \cite{2001ApJ...561..815G} to explain the optical depth fluctuations seen especially in multi-epoch pulsar observations.

Apart from understanding the mechanisms involved in the formation and the physical properties of the AU-sized structures, one of the most important questions to address first is whether the structures detected in the radio and optical observations are same or different. To investigate this we required to make observations of same sightlines. With the future facilities like SKA and TMT, this study should be possible. It would also be important to make such study in locations which are diverse in physical properties to ascertain the effects of environment on the formation of AU-sized structures.  The high resolution optical spectrometer, HROS, will be one of the first decade instruments available with TMT. When built, HROS will be capable of providing spectral resolution of R=50,000 for a 1$^{\prime\prime}$ slit or R$>$1,00,000 with an image slicer in the wavelength range of 0.31-1 micron \citep{2006SPIE.6269E..1VF, 2006SPIE.6269E..30O}. The highly efficient HROS design combined with the 10-fold increase in TMT collecting area relative to Keck will enable to observed more distant stars of a region to probe small scale structures of the foreground medium in a more systematic manner.

\section{Jets and outflows from protostars}
During the early stages of their formation, young stars drive powerful jets/outflows. They play a significant role in the evolution of a protostar as they transfer angular momentum from the young protostellar system to its environment, in the absence of which accretion cannot occur and a star cannot form. Jets are believed to drive the large scale outflows. The mechanism for the launch of jets, however, is far from certain although significant advances have been made through various numerical simulation \citep{zanni07,fendt06,ouyed97}. Most successful models of jet engines invoke magnetic fields for the launch and driving of jets. It is still not clear if the engine for jet launch is the interface between the star's magnetosphere and disk \citep[X-wind model,][]{2000prpl.conf..789S} or a wide range of disk radii \citep[disk or D-wind model,][]{2000prpl.conf..759K}. In the X-wind model, jets are launched magneto-centrifugally from the inner most parts of the disks ($r \ll 0.1$~AU) which interact strongly with stellar magnetic fields. Disk-wind model, on the other hand, explains jets as magnetocentrifugally driven from larger disk radii, typically $0.1 - 1$~AU.

Protostellar jets are observed at multiple frequencies. Observations at different frequencies trace different locations and physical processes associated with jets/outflows.  For example, CO observations at \submm/\mm~wavelengths trace the molecular gas entrained and swept-up by jets which provide a fossil record of the mass loss history of the protostar \citep{bontemps96, bt99, richer00}. Optical and near-IR forbidden lines such as [S II], [O III] and [Fe II] \citep[e.g][]{bally07, nisini09} probe fast (a few tens to few hundreds of \kms), highly collimated and partially ionised jets. Because of the large line-of-sight extinction towards protostars, these optical and IR lines are generally used to trace outer parts of the jets, farther out from the central engine, and in relatively evolved protostars with tenuous envelopes. Mid- and far-IR fine structure lines  such [O I], [Fe II], [Si II], observed with Spitzer and Herschel, suffer relatively less extinction, and they trace hot gas that is currently being shocked in jets \citep{watson15}. The spatial resolution of Spitzer and Herschel, however,  is relatively poor (10\arcsec~for [OI] at 63 \micron corresponding to $>$~1000~AU at distances to nearby star forming regions) and cannot probe the innermost parts of the jet close to where it is launched. Despite the existence of a wealth of observational studies, the nature of the primary jets from protostars is far from clear.  Several questions remain: Where and how are protostellar jets launched and accelerated? How are they collimated to such high degree? How does the jet propagation and interaction affect the surrounding medium? How do the jet properties evolve with system age? How long does jet production persist?

\subsection{`Thermal' radio jets from protostars}
 Observations at cm~wavelengths have revealed the presence of compact continuum emission centered on protostars which  is often found to be elongated roughly in the direction of the large-scale jet/outflow, indicating that the cm~emission traces the base of the jet very close to the driving source \citep[e.g.][]{anglada96, reipurth02, reipurth04}. The cm~flux density generally has a positive spectral index ($F_{\nu} \propto \nu^{\alpha}$; $\alpha \: \sim\:0.6$ for winds or radio jets) and the emission is thought to be dominated by thermal free-free radiation from ionised gas \citep{reynolds86, curiel89}.  Thus, cm~emission from low-mass protostars traces collimated and ionised jets extending to a few hundred AUs from the exciting source and corresponding to material ejected from the protostars with dynamical ages of the order of a few years or less. 

Unlike the near-infrared and optical, there have been relatively fewer studies of protostellar jets in the radio regime as the emission is relatively weak, of the order of few mJy. Most of the detection of the radio jets  have been for low-mass protostars that are relatively nearby \citep{2002RMxAA..38..169G}. In the case of massive protostars, only a handful of radio jets have been detected due to the observational difficulties encountered in studying the early phases of massive stars. Sensitive polarisation measurements at cm wavelengths using VLA of a jet GGD27-28  \citep{2010Sci...330.1209C} has shown that the magnetic field lines are parallel to the jet axis upto $\sim2500$ AU and increase in intensity towards the centre. The investigation of radio jets is of significance from massive protostars as it can help us constrain the models of massive star formation as the location and timing of launch of jet can throw light on the massive star formation scenario vis-a-vis low mass star formation.

SKA will address some very fundamental questions linked to the origin and collimation of jets by probing radio emission from the jet very close to launch point from the central exciting object. At 1 GHz, SKA should be able to observe regions as close as 25-30 AU from the central protostar at 100 pc. Although radio emission detected from jets is very weak (mJy level), a sensitive interferometer like SKA will detect radio emission from nearly all jets and outflows in the solar neighbourhood, including that from low mass protostars.

\subsection{Tracing mass accretion history from jets/outflows}
Mass accretion in young stellar objects is thought to be highly time variable and episodic \citep[e.g.][]{ken90, hartken96, evans09, dunham12}.  However, a detailed picture of the time evolution of mass accretion from early protostellar phase to late pre-main sequence  phase is still missing. This is primarily because, most of the commonly used direct observational tracers of mass accretion such as UV and optical continuum excess and emission lines of \ha, \pab, \brg fall at wavelengths $\la$~2~\micron~\citep[e.g.][]{cg98, mch98, muz98, muzerolle01}, and are heavily extinguished toward protostars which are deeply embedded in their natal core. Jets and outflows from embedded protostars, on the other hand, are more readily accessible to observations than the direct accretion tracers, particularly at \fir, mm and radio wavelengths, where the line of sight extinction can be very low.  Observations of protostellar jets and outflows at these wavelengths provide important diagnostics for the energetics of mass ejection and mass loss rates from protostars \citep[e.g.][]{bt99, richer00, hollen89, reynolds86}. Moreover, mass accretion and ejection in protostars are thought to be strongly coupled \citep{watson15}. Theoretical models of mass ejection mechanisms from protostars predict a linear relation between mass loss rate from protostars, \mout, and mass accretion rate, \macc,  onto the protostar \citep{shu94, ns94, pp92, wk93, mp05, mp08}.   Thus, observed properties of mass ejection can be used to study the mass accretion history in protostars.

Observations of molecular outflows from protostars at (sub)mm wavelengths measures time-averaged flow energetic parameters, viz., the mechanical luminosity (\lmech) and the momentum flux or outflow force (\fco). Since the observed molecular outflows are driven by protostellar jets, the measured \lmech and \fco of the outflows are the rates at which kinetic energy and momentum are injected into the flow by the jets. These are time-averaged rates over the dynamical timescale of the observed molecular flow, which is typically in the range of \eten{4}$-$\eten{5}~yr.  Thus from (sub)mm CO observations we can obtain mass loss rates from protostars averaged over \eten{4}$-$\eten{5}~yr. Observations of protostellar jets in the  [O I] line at 63~\micron provide mass loss rates averaged over the cooling timescale of $\sim$100~yr \citep[e.g.][Manoj et al. in prep]{hollen85, watson85, bea16}.  On the other hand, the ionised jets observed at radio wavelengths have dynamical timescales of $\sim$ a few yr, thus measuring the instantaneous mass loss rates. With its high sensitivity and angular resolution, SKA will resolve `thermal' jets from several hundreds of low-mass protostars in star forming regions within 500~pc from us, most of which has already been observed in (sub)mm CO lines and with \herschel \citep{manoj13}.  Thus, equipped with protostellar mass loss rates smoothed over a few yr, 100 yr and \eten{4}~yr, we will be able to study the detailed time evolution of mass ejection/accretion in protostars. Such a study will place strong constraints on the frequency, amplitude and duration of episodic accretion events in protostars during their early stages of evolution.

\subsection{Non-thermal jets from protostars}
Although the radio jets from protostars show positive spectral index, in a few cases the cm~flux density in strong radio knots is found to have negative spectral slopes, indicative of non-thermal emission \citep[e.g.][]{curiel93, wilner99}. Linearly polarised emission has been detected in some of them, confirming that the emission mechanism is non-thermal synchrotron emission \citep[e.g.][]{carrasco10}.  This provides evidence for the presence of a population of electrons accelerated to relativistic energies. It is generally thought that the electrons are  accelerated to such high velocities in strong and fast shocks \citep[e.g.][]{carrasco10}, but the exact mechanism is far from clear. Negative spectral indices at cm wavelengths are observed in only a few systems, and the polarisation measurements exist for even fewer.  Linear polarisation measurements are difficult as polarised flux density is only a fraction (typically $\sim$10\%) of the total emission and the total radio emission itself is very weak, particularly in low-mass protostars. 

With the high sensitivity offered by SKA we will be able to detect the ionised jets in protostars across a wide frequency range (few GHz to MHz) and measure the spectral indices in a large number of sources to study the incidence of  non-thermal emission.  Follow-up linear polarisation measurements of sources which show negative spectral  indices will allow us to infer the energy spectrum of the population of relativistic electrons and to study the details of particle acceleration mechanisms. In addition, such studies will also map the magnetic fields in protostellar jets, which will help address questions of collimation and acceleration of jets.

\section{Formation of high-mass stars and their effect on the surrounding ISM}

High-mass stars, with their radiative, mechanical and chemical feedback, play an important role
in the dynamical and chemical evolution of the interstellar medium (ISM) and the galaxy. 
The outpouring of UV photons and 
the associated generation of HII regions, accompanied by strong stellar winds profoundly alter the surrounding
ISM. Apart from this, massive stars evolve to become Type II Supernovae and hence 
inject energy and heavy elements to the ISM. However, the formation mechanism and the very early phases
of evolution of this mass regime is still not well understood although the basic feature of the collapse 
of a rotating cloud core is applicable to all mass ranges. The question that arises is whether the formation 
mechanism of high-mass stars (greater than $\sim8$~M$_\odot$) is just a scaled up version of the low mass regime
or the processes involved are completely different. 
The formation scenario is expected to be different for the high-mass range because the accretion timescales become larger than the protostar contraction 
timescales implying that the star `switches on' (i.e reaches the zero age main-sequence, ZAMS) while
still accreting. This invokes the `radiation-pressure' problem that inhibits further accretion onto the protostar \citep{{1974A&A....37..149K},{1987ApJ...319..850W}}
thereby questioning the formation of higher mass stars which are observationally a reality. 

The current theoretical and observational status of high mass star formation has been recently reviewed by \citet{2014prpl.conf..149T}. 
Models such as monolithic collapse with scaled up parameters like larger mass accretion rates and outflows have 
been proposed \citep{2002ApJ...569..846Y}. An alternate contender is the turbulent core model where massive stars form in 
gravitationally bound cores supported by turbulence and magnetic field \citep{2003ApJ...585..850M}. In addition, competitive accretion models have been proposed by \citet{2004MNRAS.349..735B} where small stars form via 
gravitational collapse and then grow by gravitational accretion of gas that was initially unbound to the star. 
Observationally, factors that hinder the investigation of massive stars in their infancy include rarity of sources 
(owing to fast evolutionary time scales), formation in clustered environment, large distances, complex, embedded and 
influenced environment, as well as high extinction \citep{2007ARA&A..45..481Z}. Hence, observational studies to probe the various phases 
involved in high-mass star formation and the effect they have on the surrounding ISM are of crucial importance 
in validating the proposed theories.

\subsection{Understanding the Early Phases}

In the last couple of decades, the importance of multiwavelength observations has been realised in the investigation of massive star forming regions. 
The initial studies were primarily based on far-infrared colours of IRAS \citep{{1991A&A...246..249P},{1993A&AS...98..589W},{1993A&AS..101..153P}}. 
Along the lines of the well established phases of low-mass
star formation \citep{1987ARA&A..25...23S}, recently there have been a number of attempts to carry out a similar evolutionary classification for the
higher mass counterparts. These are based on signposts such as masers, near and mid-infrared emission, jets and outflows, shocked gas as well as the presence of radio emission \citep{{2008A&A...487.1119M}, {2010ApJ...721..222B}, {2013A&A...550A..21S}, {2013A&A...556A..16G}}. 
The early phases of massive star forming sites are characterised by the presence of cold clumps detected in
millimetre continuum and infrared emission, presence of water or methanol maser and low levels of radio
continuum emission \citep{{2008A&A...487.1119M},{2013A&A...550A..21S}}. For example, \citet{2010ApJ...721..222B} have examined massive star formation by investigating a number of infrared dark clouds and have
described an observational evolutionary sequence comprising of four stages: (i) quiescent clump, (ii) clump
with signature of active star formation (maser, green fuzzy\footnote{A knot with enhanced emission in the \textit{Spitzer}-IRAC 4.5 $\mu$m band
believed to be caused by shocked H$_2$, indicative
of outflows and associated with massive young stellar objects.}, or 24~$\rm \mu m$ emission), (iii) initiation of ultracompact HII region, and (iv) diffuse red clump\footnote{A dense object within an
IRDC of parsec scale with extended, enhanced 8~$\mu$m emission.}, finally leading to the formation of a young stellar stellar cluster. 
High resolution radio continuum
mapping of objects identified in stages (ii) and (iii) holds the potential of unravelling the onset of Lyman continuum photon emission, 
providing impetus for a rigorous study of these early phases as discussed below.

In the early phases, high-mass stars are associated with small and
dense HII regions known as ultra-compact HII (UCHII) regions and the recently 
identified class of hyper-compact HII (HCHII) regions. These arise due to ionisation of the surrounding interstellar medium.
The HCHII regions (size $\le 10^4$~AU, densities $\ge 10^6$~cm$^{-3}$) display a distinct property of broad radio recombination line (RRL) profiles
 ($\ge50$~km/s) as compared to the UCHII regions \citep[$30-40$~km/s][]{2004ApJ...605..285S}. According to one school of thought, HCHII regions are
 preceded by the massive young stellar object phase \citep[MYSO,][]{2007prpl.conf..181H}. These MYSOs are luminous in infrared ($>10^4$~L$_\odot$) and 
 weak in radio as they have not yet ionised the surrounding interstellar medium and the ionised emission arises due to stellar winds from the 
star-disk system itself \citep{2011MNRAS.416..972D}. Alternate groups continue to designate this phase as earliest-stage protostellar HII regions \citep{2016ApJ...818...52T}. Two classes of such objects have been discovered: (i) those where the elongated ionised gas is perpendicular to the outflow and coincides
 with the location of a dusty disk, eg S140~IRS1 \citep{{2006ApJ...649..856H},{2013MNRAS.428..609M}}, and (ii) those where the elongated ionised gas represents a radio jet and is perpendicular to
 the disk structure such as GGD27 \citep{{1993ApJ...416..208M},{2012ApJ...752L..29C}}. It is not clear what distinguishes these two classes of MYSOs and whether they simply represent stages of an
 evolutionary chronology. Considering that the massive stars have short evolutionary timescales ($\sim10^5$~yrs), these
 protostellar phases are not expected to last long. Hence, sensitive surveys to identify and isolate these objects to make the currently limited sample
 larger, will go a long way in enhancing our understanding of the earliest phases. SKA with its multi-frequency and multi-scale capabilities should help
 address this issue.  It is expected that with the sensitivity of SKA ($\sim0.1$~$\mu$Jy at 10 GHz for 10 min integration), HCHII regions formed around stars of type B1 and earlier should be sampled throughout our Galaxy. In addition, the Core Accretion model predicts that the outflow is the
 first structure to be ionised by the protostar with a flux density of $\sim (20-200)\times{(\nu/10\,{\rm GHz})}^p$~mJy for a source at a distance of 1 kpc with
 a spectral index $p = 0.4 - 0.7$, with an apparent size that is typically $\sim500$~AU at 10 GHz \citep{2016ApJ...818...52T}. It should be possible to substantiate this with SKA using multiple frequencies given that at a distance of 1 kpc, SKA should be able to resolve spatial scales upto $\sim50$~AU at 10 GHz (corresponding to a resolution of $0.05''$ for the largest baseline diameter of 150 km). This should allow us to distinguish between competing models regarding the nature of ionised gas emission, i.e whether it is arising due to thermally evaporating flow from disks \citep{2004ApJ...614..807L} or gravitationally trapped ionised accretion flow \citep{2002ApJ...580..980K}. 

The broad RRL profiles of HCHII regions are interpreted as arising due to high densities, supersonic flows, and steep density gradients, consistent with
accretion and outflows associated with the formation of massive stars. The decrease of line width with frequency in these regions demonstrates the
 importance of pressure broadening \citep{2008ApJ...672..423K}. Multiple frequency observations of RRLs exhibit the presence of red and blue shifted gas with respect to the central location, attributed to higher density gas being traced by the high frequency lines \citep{2008ASPC..387..232L}. 
The lack of detailed kinematics across the HCHII regions is a grey area at present. In order to validate the dynamical models of ionised gas
 emission in HCHII regions, it is essential to obtain high angular resolution (one tenths to one-hundredth of an arcsec) mapping of the RRLs. Usually, the RRL line flux being an order or two lower in magnitude than the continuum flux, such a study very often becomes prohibitively expensive in time. With the
 unprecendented collecting area of SKA and large frequency coverage, it will be feasible to carry out such multi-frequency measurements at spatial scales that resolve the HCHII regions, enabling precise kinematics that appear challenging at present. For instance, the SKA1-mid bands cover nearly 200 hydrogen RRLs, from H265$\alpha$ to H78$\alpha$ and H334$\beta$ to H45$\beta$. Taking advantage of the antenna configurations, multiple frequency maps at the same scale (resolution) can be obtained to understand the motion of the ionised gas. 

An additional feature of interest that has been observed in few HCHII and UCHII regions is the variability in flux densities, also referred to as flickering.
 Observational data taken across different epochs (few years) show a change in flux density as well as morphology 
\citep{{2004ApJ...604L.105F},{2006ApJ...649..856H},{2008ApJ...674L..33G}}. This is attributed to either the variation in the flux density of the ionising source itself, or to a change in absorption of the enveloping medium such as the motion of clumps in the stellar wind. \citet{2012ApJ...758..137K} investigated this variability phenomenon using various protostellar models and showed that
 HCHII regions can dwindle in size from scales of hundreds of AU (depending on the model) to near absence during the swelling of stellar radius that
 accompanies the protostellar transition from a convective to a radiative internal structure, on timescales as short as $\sim3000$~years. On the other hand, \citet{2010ApJ...725..134P} use simulations to conclude that the intermittent shielding by dense filaments in the gravitationally unstable accretion flow
around the massive star result in highly variable UCHII regions that do not grow monotonically, but rather flicker. With the angular resolution achievable by SKA1-MID, changes in size and flux density are expected to be mapped more accurately on shorter timescales allowing us to ascertain the origin of the variability.

UCHII regions display a wide variety of radio morphologies such as cometary, bipolar, 
shell, irregular, core-halo \citep{2002ARA&A..40...27C}. However, classification based on interferometric measurements could be biased by the limited range of spatial scales that the observations are sensitive to. Larger dynamic range and higher resolution observations of an HII region could lead to an alternate morphological shape, as is evident for G29.96-0.02 (Fey et al. 1995). Therefore, a complete continuum picture of the HII region is revealed by multi-configuration observations. An improvement to the classification scheme using multi-scale radio interferometric observations was provided by De Pree et al. (2005), who recommended the removal of the core-halo morphology apart from renaming the shell morphology to shell-like. The morphology of a HII region is significant as it can provide vital clues about density inhomogeneities in the molecular cloud, the age, as well as the dynamics of the ionized and molecular gas. The thermal pressure of  ionised gas drives the dynamics in compact and large HII regions. In
the earliest stages of ionization by massive star(s), such as those
associated with early phase hypercompact and ultracompact HII regions, simulations
have shown that other dominant factors such as outflow dynamics,
rate of accretion, and gravitational instabilities shape the ionized gas distribution and evolution \citep{{2007ApJ...666..976K},{2010ApJ...725..134P},{2011IAUS..270..107K}}. The complete mapping of HII regions is also expected to eliminate problems associated with the morphological classification of HII regions at all scales. In addition, with SKA observations we foresee a clarity regarding the
hierarchical density and temperature structures that manifest as the 
 association of the large-scale diffuse emission with the  compact HII regions embedded within them 
\citep{{1999ApJ...514..232K},{2001ApJ...549..979K},{2006MNRAS.372..457S}}.

\subsection{Triggered star formation in Molecular Clouds}
It is known that significant fraction of the cluster population in a star forming region is confined to just a few rich clusters \citep{Carpenter2000}. 
As the star-forming nebula evolves, the OB stars contained within it can move far from their original birthplaces. 
Moreover, molecular clouds are clumpy and inhomogeneous on all scales. Therefore, displacement of the local gas and 
creation of non-spherical ionization fronts changes the overall structure of the GMC. Studies of individual UC HII 
regions show evidence for the evolutionary sequence of the associated nebulae \citep{2009ApJS..181..255A}. These regions can mutually interact 
and influence the future course of star formation in the entire molecular cloud. The W3/ W4/ W5 star-forming complexes
are prime examples of how this can happen. This region contains some of the richest and well-studied populations of 
deeply embedded massive stars. It presents evidence of triggered star formation, wherein new generations of stars 
have been triggered by previous generations of OB stars \citep{Oey2005}. The presence of thousands of young, low-mass stars in this 
region demonstrates that the low-mass versus high-mass star formation scenarios need not necessarily work in 
isolation \citep{Megeath2008}. The W4 region is the nearest example of a galactic superbubble powered by the winds of supernovae and 
massive OB stars \citep{West2007}. Isolated clouds seen near some GMCs could have been torn away from the parent GMC by the action of 
OB star winds. Therefore, a global view of a GMC, as opposed to studies of its  UC H II regions alone, would be able to tell us why and how these regions emerge from the molecular cloud. Another well studied object, W51, is a giant complex composed of several H II regions spread over an area of a square degree. It broadly comprises of four components, each having futher sub-components, with different properties related to star formation \citep{1998AJ....116.1856C}. Radio continuum observations of W51 have been done at different times at frequencies ranging from 151 MHz to 15 GHz at spatial resolutions of arcminutes (\cite{1995MNRAS..275.755S} and references therein). In such cases, identifying the continuum sources with, say, OH (1720 MHz) sources -- the signposts of supernova-molecular cloud interaction -- is not unambiguous.  \cite{2013ApJ...771...91B} observed W51 at 74 MHz at 88$\arcsec$ reolution and found a new non-thermal source. The wide frequency range of SKA from 50 MHz to 20 GHz offers the possibility of disentangling thermal and non-thermal components of such GMCs in a better way, and at unpreceedented resolution (7 $\arcsec$ at 110 MHz and 0.25 $\arcsec$ at 20GHz) and sensitivity ($\sim$ 1$\mu\,$Jy\,hr$^{-1/2}$).  Many W51-like interesting sources lie towards the inner Galactic plane, but are difficult to observe due to the extreme complexity of the intervening medium in their line-of-sight ; these would be accessible better using SKA.



\begin{figure*}[hbt!]
\begin{center}
\includegraphics[height=7cm,angle=-90]{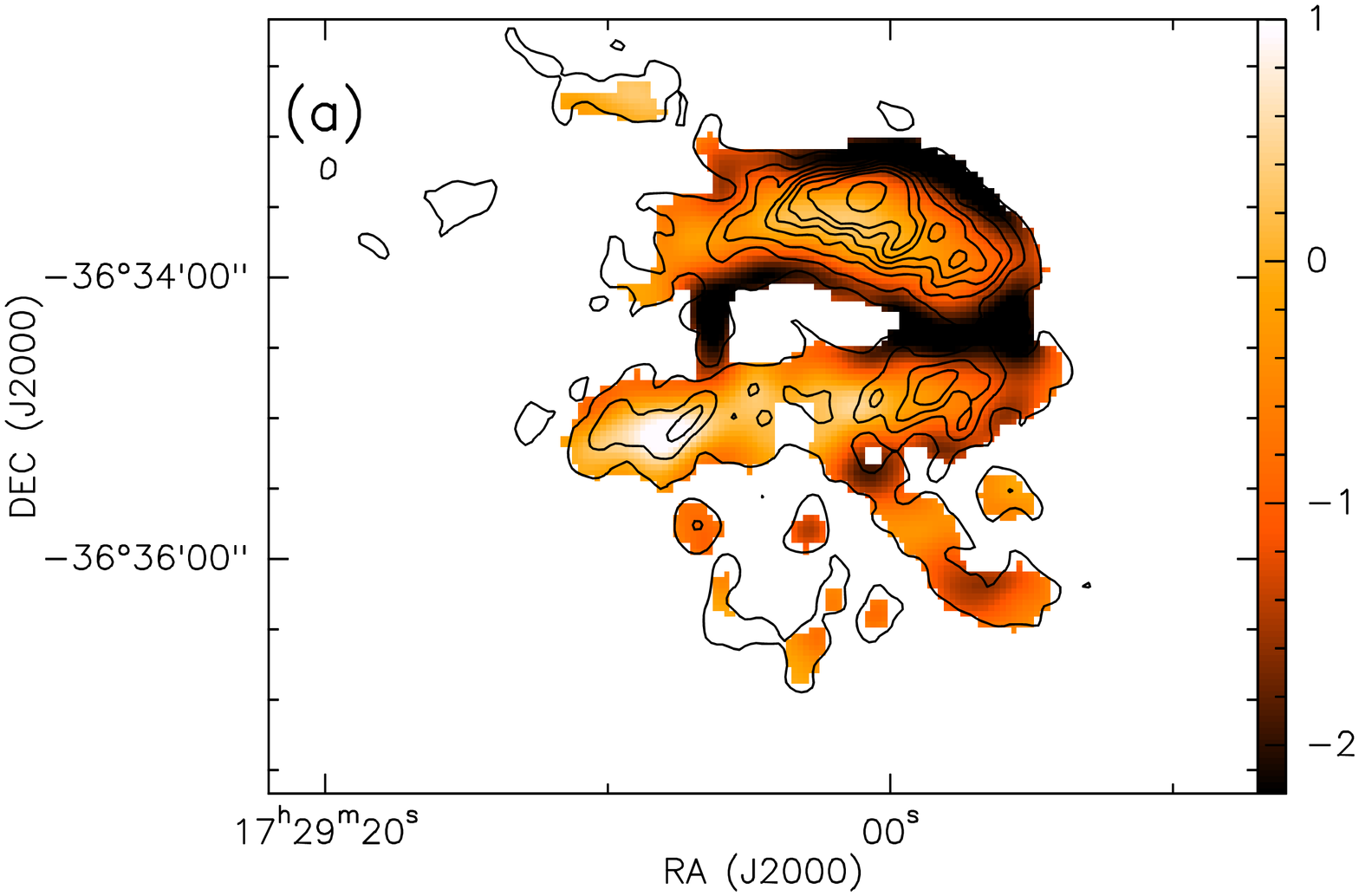}
\includegraphics[height=7cm,angle=-90]{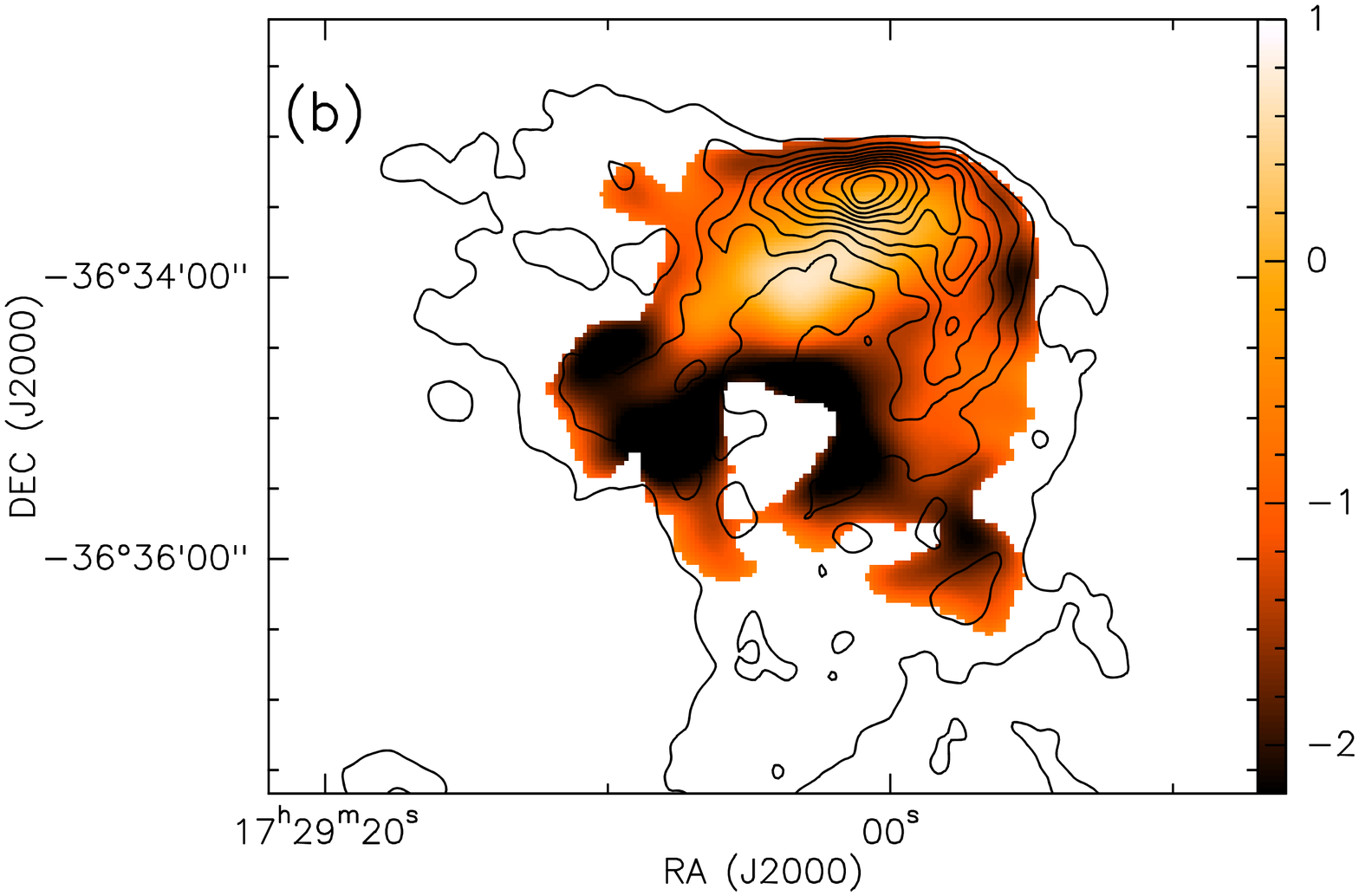}
\vskip 0.1cm
\caption{Spectral indices maps towards the cometary HII~region IRAS~17256-3631 obtained using GMRT. (a) The 1372-610 spectral index map overlaid with 1372 MHz radio continuum contours. (b) 610-325 spectral index map overlaid with 610 MHz radio continuum contours. Errors in spectral indices are $<0.5$. }
\label{si}
\end{center}
\end{figure*}


\subsection{Mid-infrared Bubbles}

An interesting manifestation of the interplay between massive stars and the 
surrounding ISM is seen in the form of bubbles which are 
observed as bright-rimmed shells prominent in the mid-infrared.
Using mid-infrared Galactic plane surveys such as GLIMPSE and MIPSGAL, a large 
number of bubbles have been identified. Catalogs of IR bubbles have been compiled by
\citet{{2006ApJ...649..759C},{2007ApJ...670..428C}} and more recently by \citet[][ - The Milky Way Project]{2012MNRAS.424.2442S}. 
These bubbles are believed to be formed around massive star(s) as is evident 
from the high correlation seen between IR bubbles and HII regions \citep{{2006ApJ...649..759C},{2007ApJ...670..428C}, {2010A&A...523A...6D}}. The observed bright-rimmed
shells are swept up gas and dust between the
ionization and the shock fronts encompassing relatively low density,
evacuated cavity around the central massive star \citep{weaver1977}. 
The mid-infrared emission is attributed to excitation of polycyclic aromatic hydrocarbon (PAH) molecules driven by the UV radiation from the central massive star(s). The formation of these bubbles is due to various feedback 
mechanisms like thermal overpressure, powerful stellar winds, radiation pressure, or 
a combination of all of them \citet{{2006ApJ...649..759C}, {2010A&A...523A...6D},{2012MNRAS.424.2442S}}.

Fragmented dust shells or clumps have been observed at the borders of several IR 
bubbles \citep{{zavagno2010},{ji2012},{liu2016}}.
Populations of young stellar objects (YSOs) are also seen towards the periphery of 
several bubbles \citep{2010A&A...523A...6D}. 
Both of the above are signposts of triggered star formation. 
The two commonly accepted models for this are the collect and
collapse \citep[(CC)][and references therein]{{weaver1977},{2010A&A...523A...6D}} and radiatively driven implosion 
\citep[(RDI)][]{1994A&A...289..559L}. The preferred mechanism (CC or RDI) and
the link with the initial star formation are still unclear. The complicated nature of this problem can be well appreciated
from studies like those by \citet{2011MNRAS.417..950H}, \citep{2002ApJ...566..302M}, \citet{2011MNRAS.414..321D}. Hence, bubbles provide an ideal database not only 
for probing high-mass star formation but also for addressing issues related to
triggered star formation.

In a recent work, \citet{das2016} have studied in detail the southern IR bubble 
S10. {\it Herschel} far-infrared images showed the presence of six clumps toward the
periphery of the bubble suggesting a fragmented shell. The derived masses of the 
clumps qualify them as potential high-mass star forming regions. Further, model based analysis revealed that these clumps harbour massive YSOs with high envelope 
accretion rates. Four of these clumps showed no radio peaks at low Giant Metrewave Radio Telescope (GMRT) frequencies of 610 and 1280~MHz and based on models are also seen to be close to the end of the 
accelerated accretion phase. Detecting the onset of ionization in the form of
resolved HCHII regions is challenging given their small sizes and large opacities
in the radio frequencies. With the achievable high resolution and sensitivity 
\citep{umana2015}, SKA has the potential to enable the detection of the 
early phases of massive star formation triggered by the expanding bubble possible
using the frequency coverage offered. 

Understanding the nature of the ionized emission in the bubble interior is
a crucial factor in validating different mechanisms proposed for its formation.
Examining the spectral index map of the bubble interior
would throw light on the nature and origin of the ionized emission in different 
zones of the bubble.
In another recent work, \citet{nandakumar2016} have discussed about non-thermal 
emission toward the periphery of the bubble CS-112. This non-thermal component
is attributed to the presence of reletivistic electrons in shocked regions of gas 
along the bubble boundary. The broad instantaneous frequency coverage of SKA would
enable obtaining spectral index maps at high resolution. 

Of recent interest is the presence of infrared arcs in the HII regions associated 
with bubbles \citep{{odorf2014},{das2016},{mackey2016}}. Some studies have
associated these with stellar wind while others have invoked the dust or bow wave 
models. High resolution radio observations achievable by SKA for locating the 
ionizing source alongwith mid-infrared observations is key to understanding the 
origin of these dust structures in the bubble interior.

\subsection{Massive Star formation studies using GMRT}

Apart from the studies mentioned in the earlier section, here we present few more examples of studies carried out by Indian Astronomers 
using GMRT. GMRT is the largest radio interferometer operating at
low frequencies: 150 - 1400~MHz. This frequency range is unique in the investigation of 
ionized emission associated with massive stars. In many cases, the frequency turn-over
occurs in this range allowing a detailed modelling of the physical conditions prevailing in HII regions. In addition, both thermal and non-thermal emission in HII regions can be examined using GMRT. 
In star formation studies, another prominent advantage of GMRT has been in the simultaneous mapping of 
compact and diffuse emission as result of the antenna distribution in a hybrid configuration. 

In the last decade, a number of studies focused on massive star forming regions have been carried
out using GMRT. Here, we discuss a few diverse results obtained. 
The radio morphology of the star forming region IRAS 19111+1048 investigated by  
\citet{2006ApJ...637..400V} shows the presence of a highly inhomogeneous ionized medium in the 
neighbourhood of an ultracompact HII region. Twenty compact sources 
including one non-thermal source were identified. The radio spectral types for majority of the 
compact sources match with the spectral type of the near-infrared counterparts. However, not all compact 
radio sources are internally excited by an embedded ZAMS star. In cases such as IRAS 17258-3637 
\citep{2014MNRAS.440.3078V} and IRAS 06055+2039 \citep{2006A&A...452..203T}, apart from the brightest 
compact source that represents the location of exciting star(s), several 
high density radio clumps have been detected that are likely to be externally ionised in a clumpy medium. 
Studies of young cluster in HII regions \citep{2015MNRAS.447.2307M} and sites of triggered
star formation associated with expanding HII regions \citep{2014A&A...566A.122S} have also been
investigated using GMRT. 

Spectral indices of diffuse emission extending upto 3~pc have been determined towards the cometary HII 
region IRAS~17256-3631 at low frequencies \citep{2016MNRAS.456.2425V}. As shown in Figure~\ref{si},
non-thermal emission is seen prominently towards the tail region that is ionisation bounded. The
lower frequency (610 - 325) spectral indices are more non-thermal
in nature (i.e. steeper negative index) when compared to the higher
frequencies, i.e. 1372 - 610. This is likely to be because thermal
contribution dominates at higher frequencies. IRAS 17256-3631
exhibits a morphology where the spectral indices are nearly flat towards
the core and relatively negative towards the diffuse envelope. It has been 
proposed that the diffuse non-thermal emission corresponds to synchrotron radiation from electrons that 
are accelerated in the region of interaction between stellar wind and ambient cloud material \citep{1996ApJ...459..193G}. It is expected that with SKA, such studies can be extended further to scales and levels that may answer as well as question our current understanding of the formation of massive stars and their interaction with the ISM.


\bibsep=0 pt
{ \footnotesize

}

\end{document}